\pdfoutput=1
\documentclass{./jfm}
\usepackage{amsmath}
\usepackage{./AMSsymb}
\usepackage{graphicx}
\usepackage{epstopdf, epsfig}
\usepackage{tikz}
\usepackage[colorlinks=true,linkcolor=black,urlcolor=blue,citecolor=blue]{hyperref}			

\shorttitle{New bounds for infinite-\Pr\ B\'enard--Marangoni convection}
\shortauthor{G. Fantuzzi, C. Nobili and A. Wynn}

\title{New bounds on the vertical heat transport for B\'enard--Marangoni convection at infinite Prandtl number}

\author{Giovanni Fantuzzi\aff{1}
  \corresp{\email{giovanni.fantuzzi10@imperial.ac.uk}},
  Camilla Nobili\aff{2}
 \and Andrew Wynn\aff{1}}

\affiliation{
	\aff{1}Department of Aeronautics, Imperial College London, London, SW7 2AZ, United Kingdom
	\aff{2}Department of Mathematics, University of Hamburg, 20146 Hamburg, Germany
}

\newcommand{\BM}{B\'enard--Marangoni }
\newcommand{\Nu}{\mbox{\it Nu}}
\newcommand{\Ma}{\mbox{\it Ma}}
\renewcommand{\Pr}{\mbox{\it Pr}}
\newcommand{\abs}[1]{\left\vert #1 \right\vert}

\newcommand{\Q}{\mathcal{Q}}

\renewcommand{\vec}[1]{\boldsymbol{#1}}
\renewcommand{\d}{{\rm  d}}
\renewcommand{\epsilon}{\varepsilon}

\definecolor{matlabblue}{RGB}{0,113,188}
\definecolor{matlabred}{RGB}{216,82,24}
\newcommand\solidrule[1][11pt]{\rule[0.5ex]{#1}{1pt}}
\newcommand\dashedrule{\mbox{%
		\solidrule[3pt]\hspace{2pt}\solidrule[3pt]\hspace{2pt}\solidrule[3pt]}}

\newcommand\dottedrule{\mbox{%
		\solidrule[1pt]\hspace{1pt}\solidrule[1pt]\hspace{1pt}\solidrule[1pt]\hspace{1pt}\solidrule[1pt]\hspace{1pt}\solidrule[1pt]\hspace{1pt}\solidrule[1pt]}}

\begin{document}

\maketitle

\begin{abstract}
We prove a new rigorous upper bound on the vertical heat transport for \BM\ convection of a two- or three-dimensional fluid layer with infinite Prandtl number. Precisely, for Marangoni number $\Ma \gg 1$ the Nusselt number $\Nu$ is bounded asymptotically by $\Nu \leq \text{const.} \times \Ma^{2/7}(\ln \Ma)^{-1/7}$. Key to our proof are a background temperature field with a hyperbolic profile near the fluid's surface and new estimates for the coupling between temperature and vertical velocity.
\end{abstract}


\section{Introduction}
\label{s:introduction}

When a layer of fluid heated from below is subject to temperature gradients along its surface, local variations in the surface tension generate a shear stress. This phenomenon, called the Marangoni effect, can set the fluid in motion when the ratio of surface tension forces to viscosity is sufficiently large. The ensuing flow, known as \BM\ convection, can produce beautiful surface patterns as famously observed by H.~\citet{Benard1901}, and is a paradigm for pattern formation. It also underpins a number of industrial processes, such as fusion welding~\citep{Debroy1995} and the growth of semiconductors~\citep{Lappa2010}. Nevertheless, \BM\ convection remains poorly understood especially when compared to its buoyancy-driven counterpart, Rayleigh--B\'enard convection. 

A fundamental open problem is to determine the vertical heat transport as a function of the thermal forcing and the material parameters of the fluid. In nondimensional terms, one is interested in how the Nusselt number $\Nu$ varies with the Marangoni number $\Ma$, which measures the relative strength of thermally-driven surface tension to viscous forces, and the Prandtl number $\Pr$, given by the ratio between the kinematic viscosity and the thermal diffusivity of the fluid.

For finite Prandtl numbers, a phenomenological argument by \citet{Pumir1996} predicts $\Nu\sim \Ma^{\frac13}$ with a Prandtl-dependent prefactor when $\Ma\gg1$ and the flow is turbulent. Two-dimensional direct numerical simulations (DNS) with stress-free boundaries at low $\Pr$ support this scaling~\citep{Boeck1998}, but no-slip boundaries in either two or three dimensions yield smaller powers of $\Ma$~\citep{Boeck2005}. Two-dimensional free-slip DNS at both high and infinite $\Pr$ also suggest a smaller exponent. Assuming steady convection rolls are stable at arbitrarily large $\Ma$, a boundary-layer scaling analysis predicts $\Nu \sim \Ma^{\frac29}$ in the infinite-$\Pr$ limit~\citep{Boeck2001}.

Rigorous results, derived directly from the governing equations without introducing unproven assumptions, are key to substantiate or rule out any of these heuristic scaling arguments. By expressing the temperature field in terms of fluctuations around a carefully chosen steady ``background'' temperature field,~\citet{Hagstrom2010} proved that $\Nu\lesssim \Ma^{\frac12}$ uniformly in $\Pr$ when this is finite,
and $\Nu\lesssim \Ma^{\frac27}$ for $\Pr=\infty$. These bounds are consistent with all aforementioned theories, but the question remains of whether they are sharp---meaning there exist convective flows that saturate them---or can be improved.

Recently, numerical optimisation of the background temperature field for $\Ma \leq 10^9$ suggested that Hagstrom \& Doering's bound for the infinite-$\Pr$ case can be improved at least by a logarithm~\citep{Fantuzzi2018a}. Precisely, the best bound available to the ``background method" for $\Ma \gg 1$ appears to be $\Nu\lesssim \Ma^{\frac27}(\ln\Ma)^{-\frac12}$, although the power of the logarithm remains uncertain due to the limited range of $\Ma$ spanned by the numerical data.
In this work, we prove analytically that logarithmic improvements to a power-law bound with exponent $2/7$ are indeed possible. Specifically, we show that 
\begin{equation}
	\label{e:Nu-bound-intro}
	\Nu \lesssim \Ma^{\frac 27}(\ln\Ma)^{-\frac 17}
	\quad {\rm  when} \quad \Ma \gg 1.
\end{equation}
We do this by combining the careful construction of an asymmetric background temperature field, inspired by the optimal profiles from~\citet{Fantuzzi2018a}, with new estimates for the coupling between temperature and vertical velocity. These differ fundamentally from the estimates that apply to infinite-\Pr\ Rayleigh--B\'enard convection~\citep{Doering2006,Whitehead2011,Whitehead2014} due to the different boundary conditions (BCs) for the velocity field.

\section{The model}
\label{s:model}

We consider a $d$-dimensional layer of fluid ($d=2$ or $3$) in a box domain with horizontal coordinates $\vec{x} \in \Pi_{i=1}^{d-1}[0,L_i]$ and vertical coordinate $z\in [0,1]$. In the infinite-$\Pr$ limit, Pearson's equations for B\'enard-Marangoni convection~\citep{Pearson1958} become
\begin{subequations}
	\begin{gather}
		\partial_t T+\vec{u}\bcdot \nabla T-\Delta T=0, \label{eq:T}\\
		\bnabla\bcdot \vec{u}=0, \label{eq:div}\\
		-\Delta \vec{u} + \nabla p=0. \label{eq:u}
	\end{gather}
\end{subequations}
Here, $\vec{u}(\vec{x},z,t)=(\vec{v}(\vec{x},z,t),w(\vec{x},z,t))$ is the velocity vector field with horizontal and vertical components $\vec{v}$ and $w$, respectively, $T(\vec{x},z,t)$ is the scalar temperature and $p(\vec{x},z,t)$ is the scalar pressure. 
We assume that all variables are periodic in the horizontal directions, while 
\begin{subequations}
	\begin{align}
		T\vert_{z=0}=0,  \qquad  \partial_z T\vert_{z=1} &= -1, \label{bc:T}\\
		\vec{u}\vert_{z=0}=0, \qquad \phantom{\partial_z} w\vert_{z=1}&=0, \label{bc:u}\\
		[\partial_z \vec{v}+\Ma\,\nabla_{\vec{x}} T]_{z=1}&=0, \label{bc:slaving}
	\end{align}
\end{subequations}
where $\nabla_{\vec{x}}$ denotes the horizontal gradient.
The steady solution $\vec{u}=0$, $T=-z$, $p={\rm const.}$ corresponds to a purely conductive state; it is globally asymptotically stable for $\Ma \leq 66.84$~\citep{Fantuzzi2017} and linearly stable for $\Ma \leq 79.61$~\citep{Pearson1958}. 

For larger Marangoni numbers convection ensues, and the velocity field can be completely slaved to the temperature.
Precisely, let $\hat{w}_{\vec{k}}$ and $\hat{T}_{\vec{k}}$ be any Fourier modes of the vertical velocity and temperature, respectively, with horizontal wavevector $\vec{k}$ of magnitude~$k$. (These are unique when $d=2$ but not when $d=3$.) One finds~\citep{Hagstrom2010}
\begin{equation}\label{eq:w-T}
	\hat{w}_{\vec{k}}(z)=-\Ma\, f_k(z)\,\hat T_{\vec{k}}(1),
\end{equation}
where, setting 
$h(x):=\frac{\sinh x}{x}$
for convenience,
\begin{equation} \label{eq:f}
	f_k(z)=\frac 12 k^2 z(z-1)\left\{\frac{h(k)h(kz)-h[k(1-z)]}{h(2k)-1}\right\}.
\end{equation}

Key to proving~\eqref{e:Nu-bound-intro} are the following new bounds for the temperature-velocity coupling in~\eqref{eq:w-T}. They are proven in Appendix~\ref{s:estimates-fk} and hold for any fixed $0 \leq \beta < 1$ and $k \geq 0$. First, for $0 \leq z \leq \beta$ we have
\begin{equation}	\label{estimate-f-small-z}
	\abs{f_k(z)}\leq \frac16\,\alpha(\beta,k)\,z^2,
	\qquad \alpha(\beta,k):= k^4\frac{ h(k)h(k\beta) }{h(2k)-1}.
\end{equation}
%
Further, for $\beta \leq z \leq 1$ we can bound
\begin{equation}
	\label{estimate-f-big-z}
	\frac{\abs{f_k(\beta)}}{1-\beta}(1-z)  \leq \abs{f_k(z)}  \leq \frac k2 (1-z) e^{-k(1-z)}.
\end{equation}

\section{Bound on the Nusselt number}
\label{s:background-method}

Denote the horizontal and long-time average of a quantity $q(\vec{x},z,t)$ by
$$\langle q \rangle(z) = \limsup_{\mathcal{T} \rightarrow \infty}\frac{1}{\mathcal{T} L_1\cdots L_{d-1}} \int_0^\mathcal{T}\!\int_0^{L_1}\cdots \int_0^{L_{d-1}} q(\vec{x},z,t)\,\d\vec{x}\,\d t.$$
%
%
Our interest is to derive a Marangoni-dependent upper bound on the Nusselt number, i.e., the ratio of the total vertical heat flux to the purely conductive one:
\begin{equation*}
	\Nu:= \frac{\int_0^1\langle wT - \partial_z T\rangle \d z}{\int_0^1\langle -\partial_z T\rangle \d z}.
\end{equation*}

To bound $\Nu$, we follow~\citet{Hagstrom2010} and write the temperature field as the sum of a steady background field $\tau(z)$, which satisfies the inhomogeneous BCs in \eqref{bc:T} but is otherwise arbitrary, and a fluctuation $\theta(\vec{x},z,t)$ satisfying
\begin{subequations}
	\begin{gather}
		T(y,z,t)=\tau(z)+\theta(y,z,t),\label{dec}\\
		\tau(0)=0, \quad \tau'(1)=-1 \label{bc:tau},\\
		\theta\vert_{z=0}=0, \quad \partial_z\theta\vert_{z=1}=0. \label{bc:theta}
	\end{gather}
\end{subequations}
Primes denote differentiation in $z$. It is shown by~\citet{Hagstrom2010} that 
\begin{equation*}
	\Nu^{-1}
	= \int_0^1 \langle \abs{\nabla\theta}^2 + 2 \tau' w \theta \rangle\, \d z - \|\tau'\|_2^2 - 2\tau(1)\,,
\end{equation*}
where $\|\bcdot\|_2$ denotes the usual $L^2$-norm.
%
At this stage, suppose that $\tau$ is chosen such that
\begin{equation}\label{e:tau-constraint}
	\Q^{\tau}\{\theta\} := \int_0^1 \langle \abs{\nabla\theta}^2 + 2 \tau' w \theta \rangle\, \d z \geq 0
\end{equation}
for all \textit{time-independent} trial fields $\theta=\theta(\vec{x},z)$ that are horizontally periodic  and satisfy~\eqref{bc:theta}, with $w = w(\vec{x},z)$ being a function of $\theta$ defined in Fourier space according to~\eqref{eq:w-T}. This can be interpreted as a nonlinear stability condition for $\tau$ \textit{as if it were} a solution to~\eqref{eq:T}--\eqref{eq:u}~\citep[see, e.g.,][]{Malkus54}. Then,
\begin{equation}\label{upper-bound}
	\Nu^{-1}
	\geq - \|\tau'\|_2^2 -2\tau(1) = 1 - \|\tau' + 1\|_2^2,
\end{equation}
where $\tau(0)=0$ is used to obtain the second equality. If the right-hand side is positive, inverting this lower bound produces a finite upper bound on $\Nu$. A background field $\tau$ is now constructed which gives~\eqref{e:Nu-bound-intro} when $\Ma \gg 1$.

\section{Proof of the main result}

The boundary condition $\tau(0)=0$ can be dropped because $\tau$ can always be shifted by a constant without affecting~\eqref{e:tau-constraint} and~\eqref{upper-bound}, which depend only on~$\tau'$. Moreover, the boundary condition $\tau'(1)=-1$ can formally be ignored because it can be enforced at the end by modifying $\tau'$ in a infinitesimally thin layer near $1$ without affecting our bound on~\Nu.
Given these observations, and motivated by the numerically optimal profiles computed by \citet[see Figure~4]{Fantuzzi2018a}, we choose
\begin{equation}
	\label{generic-profile}
	\tau'(z):=\begin{cases}
		\displaystyle-1+\left(\frac{z}{\delta}\right)^{\frac 1s} =: \eta(z) & {\rm for~} 0\leq z\leq \delta,\\
		\xi(z) &  {\rm for~} \delta\leq z\leq 1,
	\end{cases}
\end{equation}
where $\delta<\frac12$, $s>0$, and $\xi(z)$ is a non-negative function to be specified later. With $\xi(z)=0$ and $s\rightarrow 0$ this choice yields the piecewise constant profiles already studied by \citet{Hagstrom2010} and~\citet{Fantuzzi2018a}.

By expanding $\theta$ and $w$ as Fourier series in the horizontal directions, using~\eqref{eq:w-T}, and noting that $\abs{f_k(z)}=-f_k(z)$ for $0\leq z \leq 1$, it can be shown~\citep{Hagstrom2010} that the marginal stability condition~\eqref{e:tau-constraint} holds if and only if the quadratic form
\begin{equation*}
	\label{e:Qk}
	\mathcal{Q}^\tau_{ \vec{k} } \{\hat{\theta}_{\vec{k}}\} := 
	\int_0^1 \big[\,
	\vert \hat{\theta}_{\vec{k}}'(z) \vert^2 
	+ k^2 \vert \hat{\theta}_{\vec{k}}(z) \vert^2
	+ 2\Ma\,\tau'(z)\,\abs{f_k(z)}\, \hat{\theta}_{\vec{k}}(z) \hat{\theta}_{\vec{k}}(1)
	\big] \,\d z
\end{equation*}
is non-negative for all $k > 0$ and all real-valued functions $\hat{\theta}_{\vec{k}}(z)$ subject to
\begin{equation}
	\label{e:bc-spectral-constraints}
	\hat{\theta}_{\vec{k}}(0) = 0, \qquad \hat{\theta}_{\vec{k}}'(1) = 0.
\end{equation}
Since $\mathcal{Q}^\tau_{ \vec{k} } \{\hat{\theta}_{\vec{k}}\}$ is homogeneous, we may assume without loss of generality that
$\hat{\theta}_{\vec{k}}(1) \geq 0$. 

Using \eqref{generic-profile} and dropping the non-negative term $k^2\vert \hat{\theta}_{\vec{k}}(z) \vert^2$ we obtain
\begin{equation*}
\Q^{\tau}_{ \vec{k} }\{\hat{\theta}_{\vec{k}}\}
\geq \|\hat{\theta}_{\vec{k}}'\|_2^2
+2\Ma\,\hat{\theta}_{\vec{k}}(1) \!\int_0^{\delta}\eta(z)|f_k(z)|\hat{\theta}_{\vec{k}}(z)\,\d z
+2\Ma\,\hat{\theta}_{\vec{k}}(1) \!\int_{\delta}^1\xi(z)|f_k(z)|\hat{\theta}_{\vec{k}}(z)\,\d z.
\end{equation*}
The fundamental theorem of calculus, the BCs~\eqref{e:bc-spectral-constraints} and the Cauchy--Schwarz inequality imply
\begin{subequations}
	\begin{gather*}
		\hat{\theta}_{\vec{k}}(z)= \int_0^z\hat{\theta}_{\vec{k}}'(\zeta)d\zeta\leq \|\hat{\theta}_{\vec{k}}'\|_2 \sqrt z,\\
		\hat{\theta}_{\vec{k}}(z)= \hat{\theta}_{\vec{k}}(1)-\int_z^1\hat{\theta}_{\vec{k}}'(\zeta)\,d\zeta \geq \hat{\theta}_{\vec{k}}(1)-\|\hat{\theta}_{\vec{k}}'\|_2\sqrt{1-z}.
	\end{gather*}
\end{subequations}
Since the boundary value $\hat{\theta}_{\vec{k}}(1)$ and the function $\xi(z)$ are non-negative by assumption, we can use these inequalities to bound
\begin{align}
	\label{e:Qk-bound-1}
	\Q_{\vec{k}}^{\tau}\{\hat{\theta}_{\vec{k}}\}
	\geq  
	\|\hat{\theta}_{\vec{k}}'\|_2^2 
	&+2\Ma\, I_0(\xi,k) \hat{\theta}_{\vec{k}}(1)^2	
	\notag\\
	&\quad -2\Ma\,\hat{\theta}_{\vec{k}}(1)\|\hat{\theta}_{\vec{k}}'\|_2 \left[ \int_0^{\delta}\!|\eta(z) f_k(z)|\sqrt{z}\,\d z +  I_{\frac12}(\xi,k)\right],	
\end{align}
where we have introduced the notation
\begin{equation*}
	I_{\beta}(\xi,k)=\int_{\delta}^1\xi(z)|f_k(z)|\,(1-z)^{\beta}\,\d z.
\end{equation*}

Let us now estimate the terms inside the square brackets in~\eqref{e:Qk-bound-1}. For the integral over $(0,\delta)$, we use estimate \eqref{estimate-f-small-z} with $\beta=\delta$ and the definition of $\eta(z)$ from~\eqref{generic-profile} to obtain
\begin{align*}
\int_0^{\delta} \abs{\eta(z)f_k(z)}\sqrt{z}\, \d z
&\leq \int_0^{\delta} \abs{-1+\left(\frac{z}{\delta}\right)^{\frac 1s}} \, \frac{1}{6} \alpha(\delta,k) z^2 \sqrt{z}\, \d z\\
&= \frac{1}{6} \alpha(\delta,k) \int_0^{\delta} \left( z^{\frac 52} - \delta^{-\frac 1s} z^{\frac 1s + \frac 52} \right) \d z\\
&= \frac{2}{21(2+7s)} \, \alpha(\delta,k) \, \delta^{\frac 72}.
\end{align*}
To bound $I_{\frac12}(\xi,k)$, instead, we use the Cauchy--Schwarz inequality: 
\begin{align*}
I_{\frac12}(\xi,k) 
&= \int_{\delta}^1 \xi(z)|f_k(z)|\,\sqrt{1-z}\,\d z\\
&= \int_{\delta}^1 \sqrt{\xi(z)|f_k(z)|} \sqrt{ \xi(z)|f_k(z)| (1-z) }\,\d z\\
&\leq \left( \int_{\delta}^1 \xi(z)|f_k(z)|\,\d z \right)^\frac12 \left( \int_{\delta}^1 \xi(z)|f_k(z)|\,(1-z)\,\d z \right)^\frac12\\
&= \sqrt{I_0(\xi,k)} \sqrt{I_1(\xi,k)}.
\end{align*}
Substituting these two estimates into~\eqref{e:Qk-bound-1} we arrive at
\begin{align*}
\Q_{\vec{k}}^{\tau}\{\hat{\theta}_{\vec{k}}\}
\geq  
\|\hat{\theta}_{\vec{k}}'\|_2^2 
&+2\Ma\, I_0(\xi,k) \hat{\theta}_{\vec{k}}(1)^2
\notag\\
&\quad -2\Ma\,\hat{\theta}_{\vec{k}}(1)\|\hat{\theta}_{\vec{k}}'\|_2 \left[ \frac{2\,\alpha(\delta,k)\,\delta^{\frac 72} }{21(2+7s)}  +  \sqrt{I_0(\xi,k)} \sqrt{I_1(\xi,k)}\right].
\end{align*}

The right-hand side of this estmate is a quadratic form of type $ax^2 - 2bxy+ cy^2$ with $x = \|\hat{\theta}_{\vec{k}}'\|_2$ and $y = \hat{\theta}_{\vec{k}}(1)$. Quadratic forms are non-negative when their discriminant is negative, meaning $\abs{b}\leq \sqrt{ac}$, so $\Q_{\vec{k}}^{\tau}\{\hat{\theta}_{\vec{k}}\} \geq 0$ for all admissible fields $\hat{\theta}_{\vec{k}}$ if 
\begin{equation*}
\Ma \left( \frac{2\,\alpha(\delta,k)\,\delta^{\frac 72}}{21(2+7s)} +  \sqrt{I_0(\xi,k)}\sqrt{I_1(\xi,k)} \right) \leq \sqrt{2 \Ma\, I_0(\xi,k)}.
\end{equation*}
For simplicity, we rewrite this condition as
\begin{equation}
	\label{e:inequality-key}
	\frac{2\,\alpha(\delta,k)\,\delta^{\frac 72}}{21(2+7s)\sqrt{I_0(\xi,k)}} +  \sqrt{I_1(\xi,k)} \leq \sqrt{\frac{2}{\Ma}}.
\end{equation}

To prove a bound on the Nusselt number \Nu\, we require inequality~\eqref{e:inequality-key} to hold for all $k>0$. A sufficient condition for this is that $\xi(z)$ and $\delta$ be chosen such that, for some constant $c \in (0,1)$,
\begin{gather*}
	\sup_{k>0} \; \sqrt{I_{1}(\xi, k)}  \leq (1-c)\,\sqrt{\frac{2}{\Ma}},
	\\
	\sup_{k>0} \;\frac{ 2\,\alpha(\delta,k) \,\delta^\frac72}{21(2+7s)\sqrt{I_0(\xi,k)}} \leq c\,\sqrt{\frac{2}{\Ma}}.
\end{gather*}
Equivalently, after squaring both sides of each condition and rearranging,
\begin{subequations}
	\label{e:admissibility1-and-2}
	\begin{gather}
		\label{admissibility1}
		\sup_{k>0} \; I_{1}(\xi, k)  \leq \frac{2 (1-c)^2}{\Ma},\\[1ex]
		\label{admissibility2}
		\delta^7 \leq \frac{441(2+7s)^2 c^2}{2\Ma} \times \inf_{k>0} \;\frac{I_0(\xi,k)}{\alpha(\delta,k)^2}.
	\end{gather}
\end{subequations}


We will now show that~{(\ref{e:admissibility1-and-2}{\it a},{\it b})} can be satisfied by a suitable choice of~$\xi(z)$. Inspired by the numerically optimal background fields in~\citet[Figure~4]{Fantuzzi2018a} we consider
\begin{equation}
	\label{e:xi-profile}
	\xi(z) := \begin{cases}
		\displaystyle\frac{\omega\varepsilon^2}{(1-z)^{2}} &{\rm for~} 1-\gamma \leq z \leq 1-\epsilon,\\[1ex]
		0 &{\rm otherwise}.
	\end{cases}
\end{equation}
Here, $\epsilon$, $\gamma$ and $\omega$ are strictly positive parameters, to be determined as a function of the Marangoni number $\Ma$ subject to the constraint $\epsilon < \gamma \leq \frac12$.

Upon combining this choice with the upper bound on $\abs{f_k}$ in~\eqref{estimate-f-big-z} and the elementary inequality $e^{-k\varepsilon} - e^{-k\gamma} \leq 1$ we can estimate
\begin{equation*}
	I_1(\xi,k) 
	\leq \frac{k\omega \epsilon^2}{2}\int_{1-\gamma}^{1-\epsilon} e^{-k(1-z)} \,\d z 
	= \frac{\omega \epsilon^2}{2} \left( e^{-k\varepsilon} - e^{-k\gamma}\right)
	\leq \frac{\omega \epsilon^2}{2}.
\end{equation*}
This estimate holds for all $k$, so we can bound the left-hand side of~\eqref{admissibility1} from above as
\begin{equation}
\label{e:estimate-above-admissibility1}
\sup_{k>0} \; I_{1}(\xi, k) \leq \frac{\omega \epsilon^2}{2}.
\end{equation}
%
%
To estimate the right-hand side of~\eqref{admissibility2} from below, instead, observe that the lower bound on $\abs{f_k}$ in~\eqref{estimate-f-big-z} with $\beta=1-\gamma$ implies
\begin{equation*}
	I_0(\xi,k) 
	= \int_{1-\gamma}^{1-\varepsilon} \abs{f_k(z)} \frac{\omega \varepsilon^2}{(1-z)^2} \d z
	\geq \int_{1-\gamma}^{1-\varepsilon} \frac{ \omega \varepsilon^2 \abs{f_k(1-\gamma)} }{\gamma(1-z)} \d z
	= \frac{\omega \epsilon^2}{\gamma} \abs{f_k(1-\gamma)} \ln\!\left(\frac{\gamma}{\epsilon}\right).
\end{equation*}
Thus,
\begin{equation}
\label{e:estimate-below-admissibility2}
\inf_{k>0} \; \frac{I_0(\xi,k)}{\alpha(\delta,k)^2}
\geq 
\inf_{k>0} \;\frac{\omega \epsilon^2 \abs{f_k(1-\gamma)} }{\gamma \alpha(\delta,k)^2} \ln\!\left(\frac{\gamma}{\epsilon}\right).
\end{equation}

After substituting the expressions for $\abs{f_k(1-\gamma)}=-f_k(1-\gamma)$ and $\alpha(\delta,k)$ from~\eqref{eq:f} and~\eqref{estimate-f-small-z} into the right-hand side of the last inequality and rearranging, we conclude from~\eqref{e:estimate-above-admissibility1} and~\eqref{e:estimate-below-admissibility2} that conditions~\eqref{admissibility1} and~\eqref{admissibility2} hold, respectively, if
\begin{subequations}
	\label{admissibility-both}
	\begin{gather}
		\label{eps-admissibility}
		\omega \epsilon^2 \leq \frac{4 (1-c)^2}{\Ma},
		\\
		\label{del-admissibility}
		\delta^7 \leq \frac{441}{4 \Ma} (1-\gamma) \omega \epsilon^2 \left(2+7s\right)^2 c^2  \ln \left(\frac{\gamma}{\epsilon}\right) \varphi(\gamma,\delta),
	\end{gather}
\end{subequations}
where
\begin{equation*}
	\varphi(\gamma,\delta) := \inf_{k>0} \frac{ \{ h(k) h[k(1-\gamma)] - h(k\gamma) \} [ h(2k) - 1] }{k^6 h(k)^2 h(k\delta)^2}.
\end{equation*}
Observe that the right-hand side of~\eqref{del-admissibility} is strictly positive because the function ${z \mapsto h(z)}$ is increasing, so for all $\gamma,\delta \in \left[0,\tfrac12 \right]$ the quantity $\varphi(\gamma,\delta)$ satisfies
\begin{equation}
	\label{e:phi-estimates}
	0<\varphi\!\left(\tfrac12,\tfrac12\right) \leq \varphi\!\left(\gamma,\tfrac12\right) \leq \varphi(\gamma,\delta) \leq \varphi(0,0).
\end{equation}

The analysis we have just carried out shows that the background temperature field $\tau(z)$ defined through~\eqref{generic-profile} satisfies the marginal stability constraint~\eqref{e:tau-constraint} when $\xi(z)$ is as in~\eqref{e:xi-profile}, provided that~(\ref{admissibility-both}{\it a},{\it b}) hold. Let us now turn the attention to the bound on the Nusselt number produced by $\tau$. Substituting~\eqref{generic-profile} and~\eqref{e:xi-profile} into~\eqref{upper-bound} gives
\begin{equation} \label{nu-bound}
	\Nu^{-1} \geq \frac{2}{2+s}\delta 
	- \frac{\omega^2 \epsilon}{3}\left(1-\frac{\epsilon}{\gamma}\right)\left(\frac{6}{\omega} + 1 + \frac{\epsilon}{\gamma} + \frac{\epsilon^2}{\gamma^2}\right). 
\end{equation}
%

\begin{figure}
	\vspace{2ex}
	\includegraphics[scale=1]{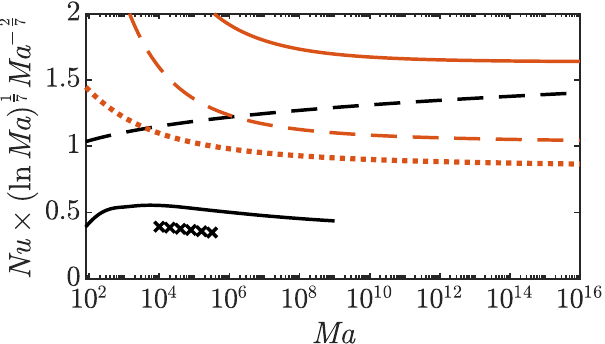} \hspace{1pt}
	\includegraphics[scale=1]{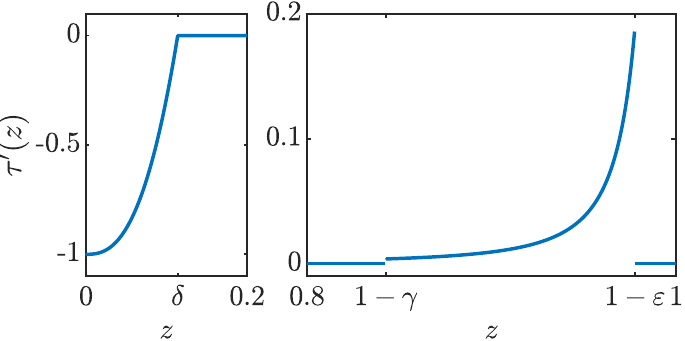}
	\\[1ex]
	\hspace*{-5pt}
	\includegraphics[scale=1]{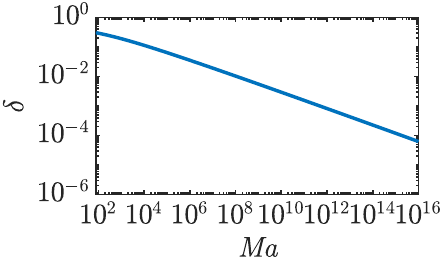}\hfill
	\includegraphics[scale=1]{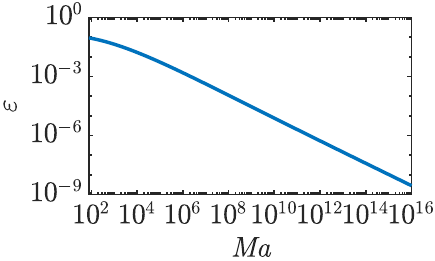}\hfill
	\includegraphics[scale=1]{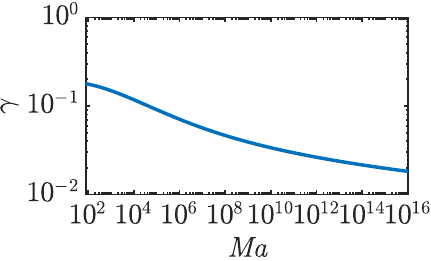}
	\\[1ex]
	\includegraphics[scale=1]{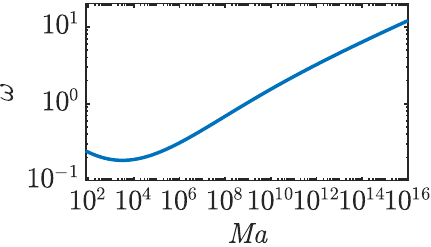}\hfill
	\includegraphics[scale=1]{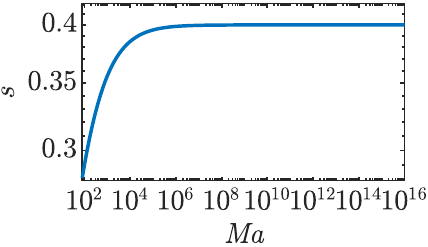}\hfill
	\includegraphics[scale=1]{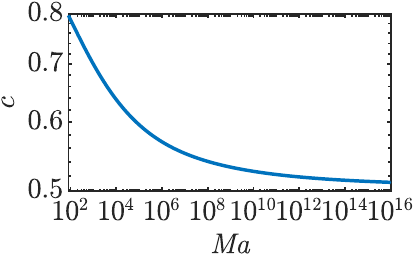}\\[-1ex]
	\begin{tikzpicture}[overlay]
	\draw node at (5.5,6.85) {(a)};
	\draw node at (7.6,8.85) {(b)};
	\draw node at (9.8,8.85) {(c)};
	\draw node at (3.9,5.15) {(d)};
	\draw node at (8.45,5.15) {(e)};
	\draw node at (13,5.15) {(f)};
	\draw node at (3.9,1.25) {(g)};
	\draw node at (8.45,1.25) {(h)};
	\draw node at (13,2.4) {(i)};
	\end{tikzpicture}
	\caption{\label{fig-numerical-bounds}
		(a)~Bounds on \Nu\
		obtained with~\eqref{nu-bound} for optimised $\delta$, $\epsilon$, $\gamma$, $s$, $\omega$, $c$ ({\color{matlabred}\dottedrule}),
		and with~\eqref{nu-bound-2} for $\epsilon=\Ma^{-1/2}$, $s=\frac25$, $c=\frac12$
		and either $\gamma = \frac12$ ({\color{matlabred}\solidrule}) or
		$\gamma = (\ln \Ma)^{-1}$ ({\color{matlabred}\dashedrule}).
		Also plotted are
		the analytical bound $\Nu \leq 0.838 \Ma^{2/7}$ by~\citet{Hagstrom2010}~(\dashedrule),
		the numerical bound by~\citet{Fantuzzi2018a}~(\solidrule), 
		and DNS data by~\citet{Boeck2001}~({\bfseries\sffamily x}).
		(b--c)~Optimised boundary layers of $\tau'$ for $\Ma=10^4$.
		(d--i) Values of $\delta$, $\epsilon$, $\gamma$, $\omega$, $s$ and $c$ that optimise \eqref{nu-bound} subject to~{(\ref{admissibility-both}{\it a},{\it b})}, $\delta < \frac12$, $\epsilon < \gamma \leq \frac12$ and $0<c<1$, as a function of~$\Ma$.
	}
\end{figure}

Maximising the right-hand side of \eqref{nu-bound} over $\delta$, $\epsilon$, $\gamma$, $s$, $\omega$ and $c$ subject to~{(\ref{admissibility-both}{\it a},{\it b})} and the constraints $\delta < \frac12$, $\epsilon < \gamma \leq \frac12$ and $0<c<1$ is hard analytically, but can be done numerically. The results, plotted in figure~\ref{fig-numerical-bounds}, 
strongly suggest that the optimal upper bound on \Nu\ provable via~\eqref{nu-bound} and~{(\ref{admissibility-both}{\it a},{\it b})} is proportional to $\Ma^\frac27 (\ln \Ma)^{-\frac17}$ as $\Ma \to \infty$, even though not all of the parameters $\delta$, $\varepsilon$, $\gamma$, $\omega$, $s$, and $c$ exhibit a simple scaling behaviour.
Optimisation of these parameters in the limit of infinite Marangoni number is also not easy and will not be pursued in this work. Instead, we prove that
\begin{equation}
\label{e:nu-bound-final-displayed}
	\Nu \lesssim 
	\Ma^{\frac27} (\ln \Ma)^{-\frac17}
	\quad {\rm as} \quad \Ma \to \infty
\end{equation}
if we set either $\gamma = \frac12$ or $\gamma = (\ln \Ma)^{-1}$ (the latter gives a better prefactor) and
\begin{subequations}
	\label{e:analytical-choices}
	\begin{gather}
		s = \frac25, \qquad
		c = \frac12, \qquad
		\epsilon = \Ma^{-\frac12}, \qquad
		\omega = 1,
		\\
		\label{e:analytical-delta-full-expression}
		\delta = \bigg[ \left( \frac{126}{5\Ma} \right)^2  (1-\gamma)  \ln \left(\gamma \Ma^{\frac12}\right) \varphi \left(\gamma,\tfrac12 \right) \bigg]^\frac17.
	\end{gather}
\end{subequations}

First, for simplicity we strengthen~\eqref{del-admissibility} by estimating $\varphi(\gamma,\delta) \geq \varphi(\gamma,\frac12)$, cf.~\eqref{e:phi-estimates}. Then, it follows from~\eqref{nu-bound} that $\delta$ should be taken as large as the resulting inequality allows. Upon insisting that
\begin{equation}\label{e:omega-epsilon-choice}
\omega \varepsilon^2 = \frac{4(1-c)^2}{\Ma}
\end{equation}
at all $\Ma$, which is the case for the optimal parameters obtained numerically, we find
\begin{equation}\label{e:delta-choice}
\delta = \left[ \frac{441}{\Ma^2} c^2(1-c)^2 (2+7s)^2 (1-\gamma)  \ln \left(\frac{\gamma}{\varepsilon}\right) \varphi \left(\gamma,\tfrac12 \right) \right]^\frac17.
\end{equation}
Substituting this expression back into~\eqref{nu-bound} and using~\eqref{e:omega-epsilon-choice} to eliminate $\omega$ yields 
\begin{equation}\label{nu-bound-2}
\Nu^{-1} \geq A(\gamma,\varepsilon,c,s) - B(\gamma,\varepsilon,c),
\end{equation}
where
\begin{gather*}
A(\gamma,\varepsilon,c,s) 
:= \frac{2}{2+s}\left[ \frac{441}{\Ma^2} c^2(1-c)^2 (2+7s)^2 (1-\gamma)  \ln \left(\frac{\gamma}{\varepsilon}\right) \varphi \left(\gamma,\tfrac12 \right) \right]^\frac17,
\\[1ex]
B(\gamma,\varepsilon,c)
:=  \frac{16(1-c)^4}{3 \Ma^2 \varepsilon^3}\left(1-\frac{\epsilon}{\gamma}\right)
	\left(\frac{3 \Ma\,\varepsilon^2}{2(1-c)^2} + 1 + \frac{\epsilon}{\gamma} + \frac{\epsilon^2}{\gamma^2}\right). 
\end{gather*}

To proceed, we make two suboptimal but simple choices. First, to simplify the dependence of $B(\gamma,\varepsilon,c)$ on $\Ma$ we set
$\varepsilon = \Ma^{-\frac12}.$
This gives $\omega=4(1-c)^2$ by~\eqref{e:omega-epsilon-choice}.
Second, motivated by our computational results we assume that $\gamma/\varepsilon = \gamma\Ma^{\frac12} \to \infty$ as $\Ma$ tends to infinity. Then, $B(\gamma,\Ma^{-\frac12},c)$ decays to zero faster than $A(\gamma,\Ma^{-\frac12},c,s)$ as $\Ma$ is raised and we conclude from~\eqref{nu-bound-2} that, asymptotically, $\Nu \leq 1/A(\gamma,\Ma^{-\frac12},c,s)$. Minimising this asymptotic bound over $s$ and $c$ simply requires maximising $A(\gamma,\Ma^{-\frac12},c,s)$. This is straightforward and yields $s=\frac25$ and $c = \frac12$, the same values approached by the optimal parameters in figure~\ref{fig-numerical-bounds}(h,i).
With these values,~\eqref{e:delta-choice} reduces to the value in~\eqref{e:analytical-delta-full-expression} and the asymptotic bound on \Nu\ becomes
\begin{equation}\label{e:optimal-bound-Nu}
\Nu \leq \frac{6}{5}\left[ \frac{126^2}{25} \frac{1}{\Ma^2} (1-\gamma)  \ln \left(\gamma \Ma^\frac12\right) \varphi \left(\gamma,\tfrac12 \right) \right]^{-\frac17}
\quad {\rm as} \quad \Ma \to \infty.
\end{equation}

Minimising this expression over $\gamma$ is not possible analytically, but is also not necessary in order to prove~\eqref{e:nu-bound-final-displayed}. For instance, simply setting $\gamma=\frac12$ gives
\begin{equation*}
	\Nu \leq
	\frac{6}{5}\left[ \frac{126^2}{100} \varphi \left(\tfrac12,\tfrac12 \right) \right]^{-\frac17} \times 
	\frac{\Ma^{\frac27}}{(\ln \Ma)^{\frac17}}
	\quad {\rm as} \quad \Ma \to \infty.
\end{equation*}
Moreover, in light of~\eqref{e:phi-estimates} the prefactor can be improved by letting $\gamma \to 0$ as $\Ma \to \infty$, which asymptotically optimises the term $(1-\gamma)\varphi(\gamma,\tfrac12)$ in~\eqref{e:optimal-bound-Nu}. The decay of $\gamma$ must be sufficiently slow to ensure that $\gamma\Ma^{\frac12} \to \infty$, as assumed above. With $\gamma = (\ln \Ma)^{-1}$, for instance,
\begin{equation*}
\Nu \leq
\frac{6}{5}\left[ \frac{126^2}{50} \varphi \left(0,\tfrac12 \right) \right]^{-\frac17} \times 
\frac{\Ma^{\frac27}}{(\ln \Ma)^{\frac17}}
\quad {\rm as} \quad \Ma \to \infty.
\end{equation*}
The exact bounds on \Nu\ obtained from~\eqref{nu-bound-2} at finite \Ma\ for~${\varepsilon=\Ma^{-\frac12}}$, $c=\frac12$, $s=\frac25$ and either $\gamma=\frac12$ or $\gamma = (\ln \Ma)^{-1}$ are plotted in figure~\ref{fig-numerical-bounds}(a).

\section{Conclusion}

In this paper we have derived a new rigorous bound for the Nusselt number in Pearson's model of \BM convection at infinite Prandtl number. Specifically, we have proven that $\Nu \lesssim \Ma^{2/7}(\ln \Ma)^{-1/7}$ at asymptotically high $\Ma$, thereby refining a pure power-law bound with exponent 2/7 by~\cite{Hagstrom2010}. The quantitative improvement on this previous result is not large for realistic values of the Marangoni number, but our logarithmic correction is significant for two reasons.

First, its proof relies on a subtle balance between the width of the bottom boundary layer of our background temperature field, which drives the asymptotic scaling of \Nu, and the stabilising effect -- with respect to the marginal stability constraint~\eqref{e:tau-constraint} -- of a thin layer near the fluid's surface where the temperature increases. Qualitatively similar layers characterise the mean vertical temperature profiles observed in DNS by~\citet[Figure 2]{Boeck2001} and their coupling underpins the phenomonological scaling theory proposed by those authors. It is therefore tempting to conjecture that the heat transport in physically realised flows indeed depends on a subtle interplay between the thermal boundary layers. In order to test this hypothesis thoroughly, it would be desirable to perform numerical simulations at higher Marangoni numbers than those considered by~\cite{Boeck2001}. Further DNS would also enable one to check if our rigorous bound is sharp and if the assumptions in Boeck \& Thess' scaling argument (most notably, the stability of simple steady convection rolls) should be revised.

Second, our result is the first upper bound proven with the background method that has a logarithmic correction with \textit{negative} exponent. This is reminiscent of scaling laws obtained for wall-bounded flows through ``mixing length'' turbulent theories~\citep[see, e.g., chapter~3 in][]{DoeringGibbon1995}. While we are not aware of any such theories being applied to \BM convection, they have historically motivated the development of rigorous upper-bounding theory in general, and the background method in particular~\citep{Doering1992}. In the future, it would be interesting to see if bounds with logarithmic corrections with negative exponent are provable for other flows, starting with extensions of the basic model considered in this work to more general types of thermal boundary conditions~\citep[e.g.,][]{Pearson1958,Fantuzzi2017}.

\vskip1ex\noindent
\textbf{Declaration of Interests.} The authors report no conflict of interest.

\bibliographystyle{./jfm}
\bibliography{references}

\appendix
\section{Estimates on $f_k(z)$}
\label{s:estimates-fk}

For the lower bound on $f_k(z)$ in~\eqref{estimate-f-big-z},
observe that the functions $h(kz)$ and $h[k(1-z)]$ are, respectively, increasing and decreasing on $[\beta,1]$ for any fixed $k>0$ . This means that the function
$\vert f_k(z)\vert (1-z)^{-1}$ increases on $[\beta,1]$, which yields the lower bound.

For the upper bound in \eqref{estimate-f-big-z}, instead, rewrite \eqref{eq:f} as
\begin{equation}
\label{eq:fk-equivalent-form-gk}
\abs{f_k(z)}=\frac{k(1-z)}{2}e^{-k(1-z)}g_k(z)
\end{equation}
with 
\begin{equation*}
g_k(z):= \frac{h(k)h(kz)-h(k(1-z))}{h(2k)-1} \,k z e^{k(1-z)}.
\end{equation*}
Differentiation gives
\begin{equation*}
g_k'(z)=\frac{e^{2k(1-z)} \ell_k(z)}{2(h(2k)-1)(1-z)^2},
\end{equation*}
with
$
\ell_k(z) := e^{-2k}\left( e^{2kz}\!\!\!-\!(1\!-\!z)^2 \right) + z^2(1\!-\!2k) + 2z(k\!-\!1).
$
Now, $\ell_k(0)=0=\ell_k(1)$ and $\ell_k'(0)> 0$. Further, $\ell_k'$ is the sum of a convex and a linear function, meaning that $\ell_k$ has at most two stationary points. Since $\ell_k'(1)=0$, there is at most one stationary point in $0 < z < 1$. Thus, both $\ell_k(z) \geq 0$ and $g_k'(z) \geq 0$ for $z \in [0,1]$. From this we conclude that
\begin{equation*}
g_k(z)\leq g_k(1)
= k \left[ h(k)^2-1 \right] \left[ h(2k)-1 \right]^{-1} \leq 1,
\end{equation*}
which, by~\eqref{eq:fk-equivalent-form-gk}, proves the upper bound in~\eqref{estimate-f-big-z}.

Finally, to show~\eqref{estimate-f-small-z}, use the definition of $h$ and the inequalities $1 \leq x\coth x\leq 1+\frac{x^2}{3},$ which are valid for $x\geq 0$, to obtain
\begin{equation*}
(z-1)\frac{h[k(1-z)]}{h(k)h(kz)} 
=   kz \left[ \coth(k) -\coth(kz) \right]
\leq  z \left( 1+\tfrac13 k^2 \right)-1.
\end{equation*}
Combining this estimate with equation \eqref{eq:f} and the identity $\abs{f_k}=-f_k$ gives
\begin{equation*}
\abs{f_k(z)} 
\leq \frac12 k^2 z \frac{h(k)h(kz)}{h(2k)-1} \left[1-z + z \left( 1+\tfrac13 k^2 \right) -1 \right]
= \frac16 k^4  \frac{h(k)h(kz)}{h(2k)-1}  z^2. 
\end{equation*}
Since $h$ is increasing, so $h(kz)\leq h(k\beta)$ for all $z \in [0,\beta]$, the upper bound~\eqref{estimate-f-small-z} follows.

\end{document}